\documentclass{aa}
\usepackage{txfonts}
\usepackage{graphicx}
\usepackage{amssymb}
\usepackage{natbib}
\bibpunct{(}{)}{;}{a}{}{,} 
\newcommand{\dd}{{\rm d}}
\begin{document}
\title{A statistical-mechanical explanation of dark matter halo properties}
\author{Dong-Biao Kang \and Ping He}
\offprints{D.-B. Kang}
\institute{Key Laboratory of Frontiers in Theoretical Physics, Institute of Theoretical Physics, Chinese Academy of Science, Beijing100190, China\\
\email{billkang@itp.ac.cn}}
\authorrunning{Kang \& He}
\titlerunning{explanation of dark matter halo properties}

\date{Received  / Accepted }
\abstract
{Cosmological $N$-body simulations have revealed many empirical relationships of dark matter halos, yet the physical origin of these halo properties still remains unclear. On the other hand, the attempts to establish the statistical mechanics for self-gravitating systems have encountered many formal difficulties, and little progress has been made for about fifty years.}
{The aim of this work is to strengthen the validity of the statistical-mechanical approach we have proposed previously to explain the dark matter halo properties.}
{By introducing an effective pressure instead of the radial pressure to construct the specific entropy, we use the entropy principle and proceed in a similar way as previously to obtain an entropy stationary equation.}
{An equation of state for equilibrated dark halos is derived from this entropy stationary equation, by which the dark halo density profiles with finite mass can be obtained. We also derive the anisotropy parameter and pseudo-phase-space density profile. All these predictions agree well with numerical simulations in the outer regions of dark halos.}
{Our work provides further support to the idea that statistical mechanics for self-gravitating systems is a viable tool for investigation.}


\maketitle

\section{Introduction}
\label{sec:intro}

The investigation of structures and properties of dark matter halos is one of the important open questions in modern cosmology. Cosmological $N$-body simulations have revealed many empirical relationships concerning dark halos, such as the density profile \citep{nfw97, moore99, einasto65}, the velocity dispersion, the anisotropy parameter, and the pseudo-phase-space density profile \citep{navarro10, ludlow10}. Yet the physical origin of these halo properties still remains unclear.

For about fifty years, many authors have attempted statistical-mechanical approaches to investigate the properties of self-gravitating systems \citep[to name a few,][]{ant62, lb67, shu78, tremaine86, sridhar87, stiavelli87, white87, tsallis88, spergel92, soker96, kull97, nakamura00, chavanis02, hjorth10}. However, these attempts have encountered many formal difficulties, and little progress has been made \citep{mo10}.

In a recent work \citep{hep10}, we employed a phenomenological entropy form of ideal gas, first proposed by \citet{white87}, to revisit this question. First, subject to the usual mass- and energy-conservation constraints, we calculated the first-order variation of the entropy, and obtained an entropy stationary equation. Then, incorporated in the Jeans equation, and by specifying some functional form for the anisotropy parameter, we solved the two equations numerically and demonstrated that the anisotropy parameter plays an important role in attaining a density profile that is finite in mass, energy, and spatial extent. If incorporated again with some empirical density profile from simulations, the theoretical predictions of the anisotropy parameter and the radial pseudo-phase-space density in the outer regions of the dark halos agree very well with the simulation data. Finally, we calculated the second-order variation, which reveals the seemingly paradoxical but actually complementary consequence that the equilibrium state of self-gravitating systems is the global minimum entropy state for the whole system, but simultaneously the local maximum entropy state for every and any finite volume element of the system. Our findings indicate that the statistical-mechanical approach should be a viable tool to account for these empirical relationships of dark halos, and may provide crucial clues to the development of the statistical mechanics of self-gravitating systems as well as other long-range interaction systems.

Despite its great success, we should point out that the specific entropy in that work is not defined in a consistent way. On the one hand, as addressed above, the anisotropy parameter is of great significance to get a finite density profile, whereas on the other hand, we had to use only the radial pressure (or velocity dispersion) for the entropy definition. Any combinations of other directions' velocity dispersions that included the anisotropy parameter, failed to produce correct equations, and hence were rejected. Besides, another shortcoming of that work should be pointed out, which is the agreements between all predictions and the simulations are restricted only to a narrow radial range of $r/r_{-2}>2$.

The aim of the current work is to remedy these defects. With the analogy of the isotropic Jeans equation, we introduce the effective pressure instead of the radial pressure to define the entropy. In a similar procedure, we derive a new entropy stationary equation, which can be easily integrated to attain the equation of state of the equilibrated dark halos, which \citet{hep10} failed to obtain with the radial pressure. It is this state equation that we employ to derive the halo density profile. Additionally, with this improved treatment, the radial range of the agreements between the predictions and simulations is increased by one order of magnitude.

This paper is organized as follows. In Sect.~\ref{sec:basic} we define the effective pressure and derive the equation of state for the self-gravitating dark matter halos. In Sect.~\ref{sec:results} we solve this equation approximately, and compare all results with the simulations of dark halos. We present the summary and conclusion in Sect.~\ref{sec:concl}.

\section{Basic theory and formulae}
\label{sec:basic}

\subsection{Isotropic Jeans equation and effective pressure}
\label{sec:hydro}

After experienced the violent relaxation process \citep{lb67}, the self-gravitating system settles into a virial equilibrium state, which is usually called the quasi-stationary state \citep{campa09}, which is described by the Jeans equation in spherical coordinates\footnote{Throughout this work, spherical symmetry is always assumed for the equilibrated dark matter halos.} as \citep{galdyn08}:
\begin{equation}
\label{eq:je}
\frac{\dd(\rho\sigma_r^2)}{\dd r} + 2 \beta\frac{\rho\sigma_r^2} {r} = - \rho \frac {\dd \Phi}{\dd r} = -\rho\frac{G M}{r^2},
\end{equation}
where $M(r) = \int^r_0 4\pi{r^2}\rho\dd r$, $\Phi$ is the gravitational potential, $\beta = 1 - \sigma^2_t/{2\sigma^2_r}$ is the anisotropy parameter, and $\sigma^2_t$ and $\sigma^2_r$ are the tangential and radial velocity dispersion, respectively.

For $\beta=0$, i.e. the isotropic case, the Jeans equation reduces to the following,
\begin{equation}
\label{eq:hydroe}
\frac{\dd p_r}{\dd r}=-\rho\frac{\dd\Phi}{\dd r},
\end{equation}
where $p_r=\rho\sigma_r^2$ is the radial pressure. Equation~(\ref{eq:hydroe}) resembles the hydrostatic equilibrium equation of an ordinary fluid, $\nabla p=-\rho\nabla\Phi$, in that the radial pressure gradient, the buoyancy, resists the gravitational force of the whole system. Enlightened by this similarity, we generalize Eq.~(\ref{eq:hydroe}) to the anisotropic case by defining the effective pressure $P$ through:
\begin{equation}
\label{eq:peff}
\frac{\dd P} {\dd r} = \frac{\dd p_r}{\dd r} + 2\beta \frac{p_r}{r}.
\end{equation}
Then, the Jeans equation, Eq.~(\ref{eq:je}), can be re-written in terms of the effective pressure $P$ as
\begin{equation}
\label{eq:pr}
\frac{\dd P}{\dd r} = - \rho\frac{\dd \Phi}{\dd r} = -\rho\frac{G M}{r^2}.
\end{equation}
Below we will use the effective pressure to construct the specific entropy form.

\subsection{Equation of state}
\label{sec:eos}

As mentioned in the introduction, in \citet{hep10} we employed the entropy principle and derived an entropy stationary equation by introducing the specific entropy,
\begin{equation}
\label{eq:ss}
s=\ln(p^{3/2}_r\rho^{-5/2}) = - \ln(\rho/\sigma^3_r).
\end{equation}
This entropy form was first used by \citet{white87} for the isotropic self-bounded gas sphere, but we assumed it can also be applied to the velocity-anisotropic case. This is not a self-consistent treatment, and also it's hard to generalize to include the anisotropy parameter.

Based on the above similarity between Eqs.~(\ref{eq:hydroe}) and (\ref{eq:pr}), we replace $p_r$ in Eq.~(\ref{eq:ss}) by the effective pressure $P$ to make the entropy more appropriate for the anisotropic case, so that the total entropy is
\begin{equation}
\label{eq:st}
S_{\rm t} = \int_0^{\infty}4 \pi r^2 \rho s' \dd r = \int_0^{\infty}4 \pi r^2 \rho\ln(P^{3/2} \rho^{-5/2})\dd r,
\end{equation}
where $s' = \ln(P^{3/2}\rho^{-5/2})$. Subject to the constraints of mass and energy conservation, we calculate the first-order variation of the entropy in a similar process as the one in \citet{hep10}, to obtain the new entropy stationary equation as
\begin{equation}
\label{eq:sse1}
\frac{3}{2}\frac{\dd \ln P}{\dd r} - \frac{5}{2}\frac{\dd \ln\rho}{\dd r} = -\lambda \frac{\dd P}{\rho \dd r},
\end{equation}
where $\lambda$ is the Lagrangian multiplier corresponding to the energy conservation. The left-hand side of this equation is formally the same as that of \citet{hep10}, except that $p_r$ is replaced by the effective pressure $P$.

Equation~(\ref{eq:sse1}) can be readily transformed into
\begin{equation}
\label{eq:sse2}
\frac{\dd \rho}{\dd P} = \frac{3}{5}\frac{\rho}{P} + \frac{2}{5}\lambda,
\end{equation}
and can be directly solved as
\begin{equation}
\label{eq:ees}
\rho = \lambda P + \mu P^{3/5},
\end{equation}
where $\mu$ is an integration constant. $\mu$ and $\lambda$ are related to the total mass and energy of the dark halo, that is, they can be specified by the total mass $M$, and total energy $E$, as $\mu=\mu(M, E)$ and $\lambda=\lambda(M,E)$, and vice versa.

Equation~(\ref{eq:ees}) describes the relationship between $\rho$ and $P$, which is exactly the equation of state of the equilibrated system, but $P$ is also related to $\rho$ through the differential equation, Eq.~(\ref{eq:pr}), thus Eq.~(\ref{eq:ees}) cannot directly provide us with the dark halo density profile. Below we will explore how to obtain the density profile from this state equation.

\section{Results}
\label{sec:results}

From Eq.~(\ref{eq:ees}) we can see that $\rho$ scales with $P$ as $\rho \sim \lambda P$ at the center of the dark halo, but $\rho \sim \mu P^{3/5}$ at the outskirts, since both $\rho$ and $P$ are large at the center, but small at the outskirts of the dark halos. This suggests that, it would be advantageous to analyze the approximate solutions of these two cases, before we discuss the general solution of Eq.~(\ref{eq:ees}).

\begin{figure}
\resizebox{\hsize}{!}{\includegraphics{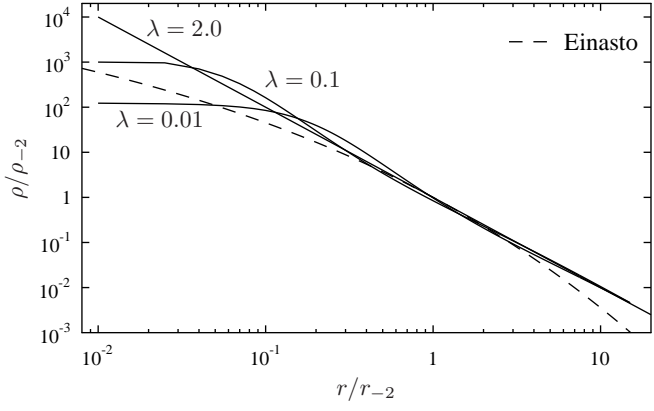}}
\caption{Isothermal density profiles. The inner slope of the density profile is dependent on different $\lambda$. The Einasto profile, $\ln(\rho/ \rho_{-2}) = -2/ \alpha ((r/r_{-2})^{\alpha}-1)$, with $\alpha \approx 0.17$ \citep[see][]{navarro10}, is also indicated for comparison.}
\label{fig1}
\end{figure}

\subsection{$\lambda\neq0, \mu=0$}
\label{sec:resub1}

As explained above, this case is the approximation of Eq.~(\ref{eq:ees}) at the halo center. If $\lambda$ is renamed as $\lambda = 1/\sigma^2_{\lambda}$ (we will see $\lambda$ must be positive in the following)\footnote{$\sigma^2_{\lambda}$ has the same dimension as the velocity dispersions $\sigma^2_r$, or $\sigma^2_t$, but they are completely different in physical meanings.}, then with $\mu=0$, Eq.~(\ref{eq:ees}) is re-expressed as
\begin{equation}
\label{eq:ise}
P=\rho\sigma^2_{\lambda}.
\end{equation}
Incorporating Eq.~(\ref{eq:pr}), and differentiating both sides of the equation with respect to $r$, we have
\begin{equation}
\label{eq:ije}
\sigma^2_{\lambda}\frac{\dd\rho}{\dd r} = -\rho\frac{\dd\Phi}{\dd r},
\end{equation}
with the solution as $\Phi=-\sigma^2_{\lambda}\ln\rho$, which resembles the equation that the isothermal gas satisfies \citep[see][p.303]{galdyn08}. Differentiating Eq.~(\ref{eq:ije}) again, we have
\begin{equation}
\label{eq:idp}
\frac{1}{r^2}\frac{\dd}{\dd r}(r^2\frac{\dd\ln\rho}{\dd r}) = -4\pi G \lambda \rho.
\end{equation}
We show the solutions in Fig.~\ref{fig1}, with the case of $\lambda = 2$ corresponding to the singular isothermal sphere\footnote{For the characteristic density $\rho_{-2}$ and scale $r_{-2}$, we set $4\pi G = \rho_{-2}=r_{-2}=1$.}, $\rho\sim r^{-2}$. We can also see that $\lambda$ with $\lambda < 2$ bends the singular isothermal solution toward a centrally-cored density profile, and the size of the core depends on the value of $\lambda$, but the slope at the outskirts of the dark halos remains unchanged.

\begin{figure}
\resizebox{\hsize}{!}{\includegraphics{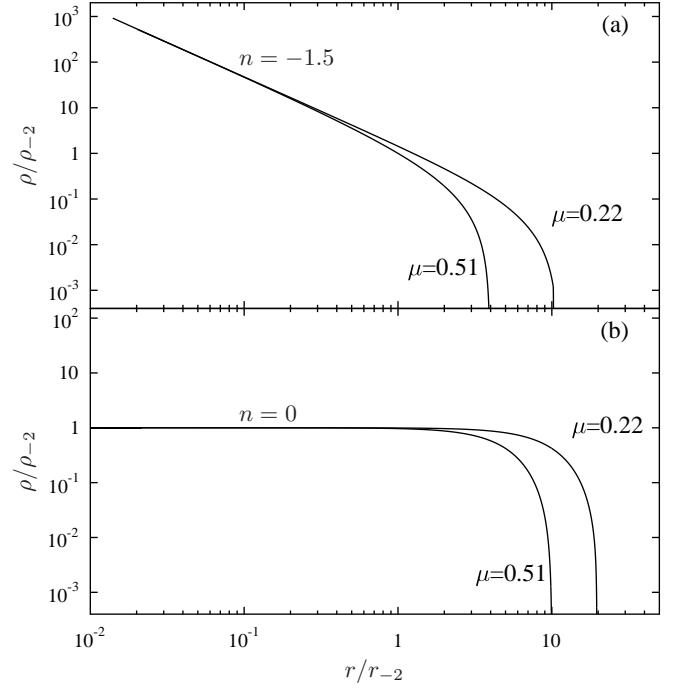}}
\caption{Polytropic density profiles. Panels (a) and (b) correspond to the two approximate inner power-law solutions, $\rho\propto{r^n}$, with the power index as $n=-1.5$ and $n=0$, respectively. All density profiles are self-truncated, with the truncation radii varying with different $\mu$.}
\label{fig2}
\end{figure}

\subsection{$\lambda=0,\mu\neq0$}
\label{sec:resub2}

In this case, the equation of state, Eq.~(\ref{eq:ees}), is reduced to
\begin{equation}
\label{eq:ose}
P = \mu^{-5/3} \rho^{5/3},
\end{equation}
which resembles a polytropic process of a common gas. Differentiating this equation with respect to $r$, and incorporating Eq.~(\ref{eq:pr}), we get the following differential equation
\begin{equation}
\label{eq:odp}
2\frac{\dd\rho}{\dd r} + \left(\frac{\dd^2\rho}{\dd r^2} - \frac{1}{3\rho} (\frac{\dd\rho}{\dd r})^2\right) r =-\frac{20\pi G}{3} \mu^{5/3}\rho^{4/3} r.
\end{equation}
As analyzed previously, this equation is just an approximation of Eq.~(\ref{eq:ees}) at the outskirts of the dark halos. As $r$ reaches zero, we can see that the right-hand side of the equation also reaches zero, so that we can approximate the solution with a power-law form, $\rho\sim r^n$, with the power-index $n=0, -1.5$. Then starting from the two power-law inner solutions at sufficient small $r$, we numerically evaluate Eq.~(\ref{eq:odp}) with different $\mu$, and show all results in Fig.~\ref{fig2}. Since Eq.~(\ref{eq:odp}) is only valid at the outskirts of dark halos, we can see that regardless of the non-uniqueness of the inner slopes at small radii, all resulting density profiles behave as if they were truncated by the $\mu$-term of this equation, so that both the mass and the energy of the dark halos are not infinite.

\begin{figure}
\resizebox{\hsize}{!}{\includegraphics{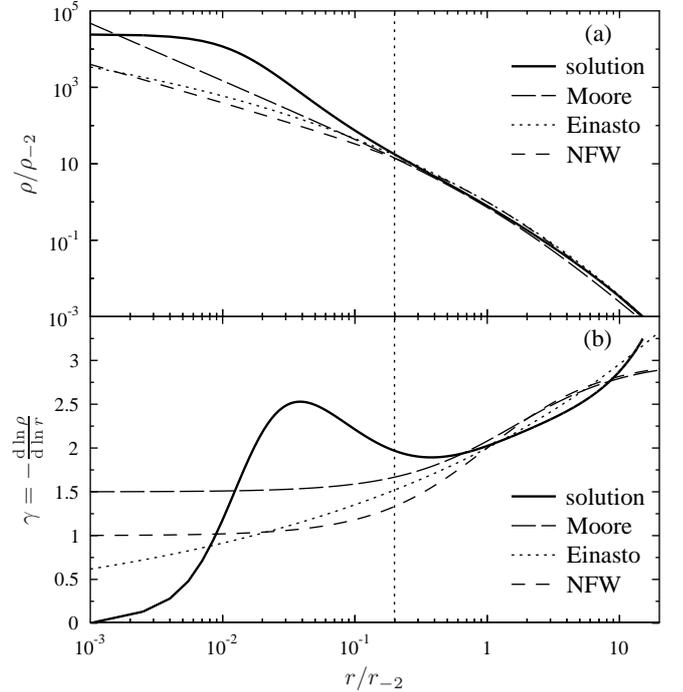}}
\caption{(a) The density profiles and (b) the logarithmic slopes corresponding to the density profiles in Panel (a). The profile and the slope derived from Eq.~(\ref{eq:cdp}) are indicated as ``solution'' in the figure, with $\lambda = 2.27$, $\mu = 0.36$. The major fitting formulae of the density profile, NFW, Moore, and Einasto profile, are also shown for comparison.}
\label{fig3}
\end{figure}

\subsection{$\lambda\neq0,\mu\neq0$}
\label{sec:resub3}

With the heuristic results from the above two subsections, we explore the complete solution of this general case. It may not be easy to solve Eqs.~(\ref{eq:pr}) and (\ref{eq:ees}) exactly, so we try to find the approximate solution to the density profile. We treat Eq.~(\ref{eq:ise}) as a first approximation and substitute it into the derivative of the state equation, Eq.~(\ref{eq:ees}), then we have
\begin{equation}
\label{eq:cdp}
\frac{\dd\rho}{\dd r} = -\rho\frac{\dd\Phi} {\dd r} (\frac{1}{\sigma^2_{\lambda}} + \frac{3\mu} {5\sigma^{4/5}_{\lambda} \rho^{2/5}}),
\end{equation}
in which again $\sigma^2_{\lambda}$ is the renaming of $\lambda$, as $\lambda = 1 / \sigma^2_{\lambda}$. We numerically evaluate this equation, and the resulting density profile is exhibited in Fig.~\ref{fig3}a. We also indicate in Fig.~\ref{fig3}b the logarithmic slope of the density profile, i.e., $\gamma = -\dd\ln\rho/\dd\ln r$. In the inner region $\rho$ indeed behaves as the isothermal solution of Eq.~(\ref{eq:idp}). In the outer region, the density profile does not decline as fast as the polytropic solution of Eq.~(\ref{eq:odp}) because we only consider the first approximation, but it is still obvious that the whole density profile is roughly the superposition of the above two subsections' solutions.

We can see that with an appropriate value of $\lambda$ and $\mu$, both the resulting density profile and its log-slope agree very well with the fitting formulae at the radii $r/r_{-2}\geq0.2$, the right part of the vertical dotted line in Fig.~\ref{fig3}. At small radii with $r/r_{-2}<0.2$, the solution is qualitatively accepted, but the agreement is not as good as the case at large radii.

\begin{figure}
\resizebox{\hsize}{!}{\includegraphics{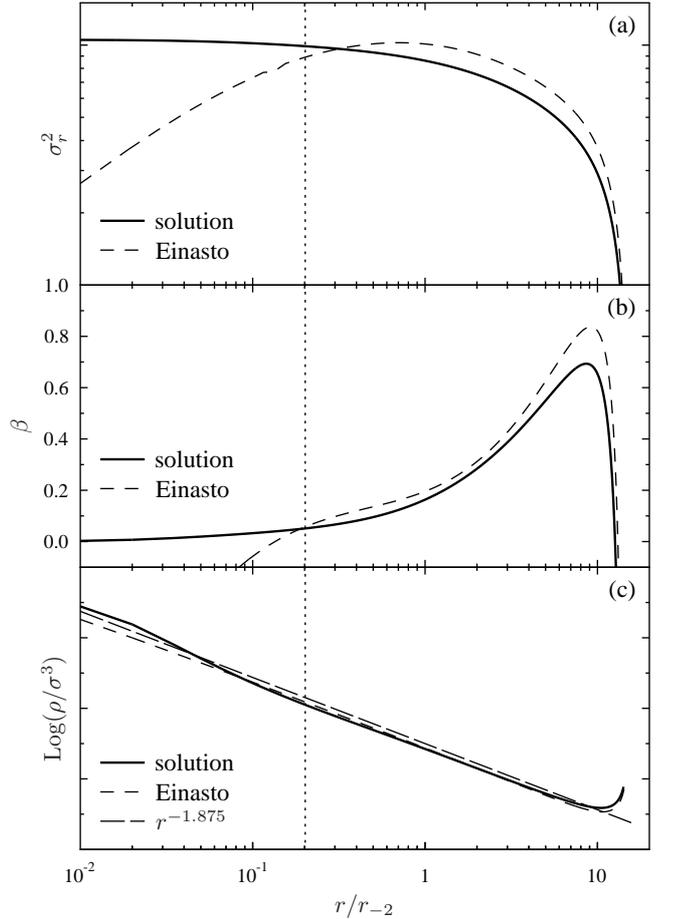}}
\caption{The radial velocity dispersion $\sigma^2_r(r)$, the anisotropy parameter $\beta(r)$, and the pseudo-phase-space density $Q(r)$ are shown in panels (a), (b), (c), respectively. The results based on the resulting density profile in Sect. \ref{sec:resub3} are indicated as ``solution'' in the figure. The results based on the Einasto profile are also shown for comparison. The vertical scales of panels (a) and (c) are arbitrary.}
\label{fig4}
\end{figure}

\subsection{$\sigma^2_r$, $\beta$, {\rm and} $\rho/\sigma^3_r$ profiles}
\label{sec:resub4}

If the density profile $\rho(r)$ is determined, we can obtain the effective pressure $P$ by integrating Eq.~(\ref{eq:pr}). Numerical simulations indicate that the anisotropy parameter, $\beta$, is usually a small number \citep{navarro10}. From Eq.~({\ref{eq:peff}}) we can see that when $\beta$ is small, we can approximate the effective pressure $P$ as
\begin{equation}
\label{eq:apr}
P \approx p_r = \rho\sigma^2_r,
\end{equation}
by which $\sigma^2_r$ can be expressed as $\sigma^2_r \approx P/\rho$.

From Eqs.~(\ref{eq:je}) and (\ref{eq:cdp}) we obtain
\begin{equation}
\label{eq:aje}
\frac{\dd(\rho\sigma^2_r)} {\dd r} + 2\beta\frac{(\rho\sigma^2_r)}{r} = \frac{\sigma^2_{\lambda}}{(1 + 3\mu\sigma^{6/5}_{\lambda}/5 \rho^{2/5})}\frac{\dd\rho}{\dd r}.
\end{equation}
Then if $\sigma^2_r$, $\rho$, and hence the logarithmic slope of the density profile  $\gamma = -\dd\ln\rho/ \dd\ln r$ are given, $\beta$ can be expressed as
\begin{equation}
\label{eq:beta}
2\beta=\left(1-\frac{\sigma^2_{\lambda}}{\sigma^2_r}{\big /}(1 + \frac{3}{5}\frac{\mu \sigma^{6/5}_{\lambda}}{\rho^{2/5}})\right) \gamma - \frac{\dd\ln\sigma^2_r} {\dd\ln r} = a(r)\gamma - b(r),
\end{equation}
where $a(r) = 1 - \sigma^2_{\lambda}/\sigma^2_r(1 + 3\mu \sigma^{6/5}_{\lambda}/5\rho^{2/5})$, $b(r) = \dd \ln \sigma^2_r/ \dd\ln r$.

We present a simple analysis of what Eq.~(\ref{eq:beta}) implies. In the inner region of dark halos $\rho$ is large, the term including $\mu$ can be neglected, and from the simulation results \citep{navarro10} we know both $a(r)$ and $b(r)$ vary slowly with $r$, so it seems that there is an approximately but not strictly linear relationship between $\beta$ and $\gamma$. The existence of this linear relation has been observed in both \citet{hansen06} and the inner region of \citet{navarro10}. Because this analysis will not be valid for regions at larger radii, the linear relationship should not exist for the whole region. Indeed, the latest simulations by \citet{navarro10} and \citet{ludlow10} show that $\beta(r)$ is not the monotonic function of $r$, that is, haloes are nearly isotropic near the center, radially biased to the maximum at some radius and approximately isotropic again in the outskirts, which is against the earlier monotonic result of \citet{hansen06}.

In the inner region, since $a<1$ and $b>0$, so from Eq.~(\ref{eq:beta}) we must have $\gamma > 2\beta$, and in the outer region, since $\gamma>2$ and $\beta<1$, we always have $\gamma > 2\beta$. So we have $\gamma > 2\beta$ in the whole region, which is consistent with the result of \citet{an05} and \citet{ciotti10}.

Next, we analyze the behavior of the pseudo-phase-space density $Q(r) \equiv \rho / \sigma^3_r$ (or $\rho/\sigma^3_{\rm tot}$). From Eq.~(\ref{eq:beta}) we have
\begin{equation}
\label{eq:ppsd}
\frac{\dd\ln Q}{\dd\ln r} = \frac{\dd\ln\rho}{\dd\ln r} - \frac{3}{2} \frac{\dd\ln \sigma_r^2}{\dd\ln r} = -\gamma - \frac{3b}{2} = -\frac{2\beta + b}{a} - \frac{3b}{2}.
\end{equation}
In the inner region we know from \citet{navarro10} that $\beta$, $a$, and $b$ all vary slowly with $r$, so that $\dd\ln Q/\dd\ln r$ roughly seems to be a constant. If we take \citet{hansen06}'s result as a reference, $a\approx0.4$, $b\approx 0.32$, and assume $\beta \approx 0.12$ on the average, then $\dd\ln Q/\dd\ln r \approx -1.88$, very close to the spherical secondary-infall similarity solution \citep{bertschinger85}.

With the density profile derived in the previous subsection (see Fig.~\ref{fig3}a), and by Eq.~(\ref{eq:apr}), we calculate the approximate radial velocity dispersion, $\sigma^2_r$, which is shown in Fig.~\ref{fig4}a. The velocity dispersion calculated with Einasto profile is also indicated in the figure for comparison. We can see that both the two results roughly agree with the simulation results of \citet{navarro10} when $r/r_{-2}>0.2$, as is consistent with the accuracy of the resulting density profile of Sect.~\ref{sec:resub3}. The discrepancy is significant when $r/r_{-2}<0.2$, which may suggest that there may be still some crucial physics that we have failed to capture in the current formulation.

Given the density profile $\rho(r)$, and the velocity dispersion $\sigma^2_r(r)$, and with the two parameters $\lambda$ (or $\sigma^2_{\lambda}$) and $\mu$ being the same as we computed the density profile, we can also obtain $\beta(r)$ via Eq.~(\ref{eq:beta}) as well as the pseudo-phase-space density $Q(r)$. The results are exhibited in Figs.~\ref{fig4}b and \ref{fig4}c, respectively. The corresponding results derived by using the Einasto profile are also indicated for comparison. From Fig.~\ref{fig4}b, we notice that the non-monotonicity of $\beta(r)$ is once again reproduced in our current work, which agrees well with the simulation results of \citet{navarro10} and \citet{ludlow10}. However, we find that both the position and the height of the peak of the $\beta(r)$ profile deviate slightly from the simulations. We speculate that this discrepancy may be originated in the approximation we adopted for Eqs.~(\ref{eq:cdp}) -- (\ref{eq:aje}). The result based on the Einasto profile cannot be accepted for $r/r_{-2}<0.2$, as is consistent with the previous results for both $\rho$ and $\sigma^2_r$.

As expected, the resulting $Q(r)$ profile closely follows a power law, $r^{-1.875}$, near the center of the dark halo. However, the predictions do show a manifest curve-up deviation from this power law at the outskirts of dark halos, which has been observed by the simulation result of \citet{ludlow10}, and has already been reproduced by our previous study \citep{hep10}.

\section{Summary and conclusion}
\label{sec:concl}

Cosmological $N$-body simulations have revealed many empirical relationships concerning dark halos. Up to now, however, there are no robust physical origins for these empirical relationships. In this work, we aim to provide a unified statistical-mechanical explanation to the relationships concerning the matter density, anisotropy parameter and the pseudo-phase-space density profile. The main steps of this work are similar to those of \citet{hep10}, i.e., by using the entropy principle with mass- and energy-conservation as constraints to derive the entropy stationary equation, Eq.~(\ref{eq:sse1}). Different from \citet{hep10}, however, based on the analogy with the isotropic Jeans equation, we phenomenologically introduce the effective pressure $P$ instead of the radial pressure $p_r$ to construct the specific entropy. The effective pressure is a key quantity in this work and proves to be more powerful than $p_r$.

The entropy stationary equation can be easily integrated to attain Eq.~(\ref{eq:ees}), the equation of state of the equilibrated dark halos, which \citet{hep10} failed to attain with the radial pressure $p_r$. It is this equation of state that we employ to derive the halo density profile $\rho(r)$. Notice that this density profile has nothing to do with $\beta$.

First, we study the approximate solutions of Eq.~(\ref{eq:ees}) in two asymptotic cases, i.e. the inner and outer region of the dark halos. The inner solution is the isothermal density profile, with the inner logarithmic slope varying with different parameter $\lambda$. The outer solution is the polytropic density profile, self-truncated at sufficiently large radii with a different parameter $\mu$, so that the density profile has finite mass, energy and extent. The approximate complete solution can be roughly regarded as the superposition of the two asymptotic solutions. By choosing appropriate values of $\lambda$ and $\mu$, we can reproduce both the density profile and its logarithmic slope quite well in the region $r/r_{-2}>0.2$. In the inner region of $r/r_{-2}<0.2$, the solution seems to be qualitatively accepted, but the agreement between the solution and the fitting formulae is not quantitatively well. We speculate that there may be still some crucial physics that we have failed to capture in the current formulation.

Then, with the resulting density profile, we compute the radial velocity dispersion $\sigma^2_r(r)$, and then $\beta(r)$ and the pseudo-phase-space density $Q(r)$ profile. We find that all these predictions agree quite well with the corresponding results of the latest simulations of \citet{navarro10} and \citet{ludlow10} in the region of $r/r_{-2}>0.2$. Despite the slight discrepancies from the simulation results in the inner regions, these agreements at the outer regions of dark halos strengthen the conclusions of \citet{hep10}, indicating once again the great success of the statistical-mechanical approaches for the self-gravitating systems.

\begin{acknowledgements}

D.B.K. is very grateful for the comments and suggestions of the anonymous referee. This work is supported by the National Basic Research Programm of China, NO:2010CB832805.

\end{acknowledgements}


\end{document}